%% file: IEEE-conference-template-062824.tex
\documentclass[conference]{IEEEtran}
\IEEEoverridecommandlockouts

\usepackage{cite}
\usepackage{amsmath,amssymb,amsfonts}
\usepackage{algorithmic}
\usepackage{graphicx}
\usepackage{textcomp}
\usepackage{xcolor}

\def\BibTeX{{\rm B\kern-.05em{\sc i\kern-.025em b}\kern-.08em
    T\kern-.1667em\lower.7ex\hbox{E}\kern-.125emX}}

\usepackage{tikz}
\usepackage{hyperref}
\usepackage{geometry}
\usepackage{enumitem}
\usepackage{caption}

\begin{document}

\title{AI Work Quantization Model: Closed-System AI Computational Effort Metric}

\author{
    \IEEEauthorblockN{Aasish Kumar Sharma, Michael Bidollahkhani, Julian Martin Kunkel}
    \IEEEauthorblockA{\textit{Faculty of Mathematics and Computer Science, GWDG} \\
    \textit{Georg-August-Universität Göttingen}\\
    Göttingen, Germany \\
    aasish-kumar.sharma@gwdg.de, michael.bkhani@uni-goettingen.de, julian.kunkel@gwdg.de}
}


\maketitle

\begin{abstract}
The rapid adoption of AI-driven automation in IoT environments, particularly in smart cities and industrial systems, necessitates a standardized approach to quantify AI’s computational workload. Existing methodologies lack a consistent framework for measuring AI computational effort across diverse architectures, posing challenges in fair taxation models and energy-aware workload assessments. This study introduces the Closed-System AI Computational Effort Metric, a theoretical framework that quantifies real-time computational effort by incorporating input/output complexity, execution dynamics, and hardware-specific performance factors. The model ensures comparability between AI workloads across traditional CPUs and modern GPU/TPU accelerators, facilitating standardized performance evaluations.

Additionally, we propose an energy-aware extension to assess AI’s environmental impact, enabling sustainability-focused AI optimizations and equitable taxation models. Our findings establish a direct correlation between AI workload and human productivity, where 5 AI Workload Units equate to approximately 60–72 hours of human labor—exceeding a full-time workweek. By systematically linking AI computational effort to human labor, this framework enhances the understanding of AI’s role in workforce automation, industrial efficiency, and sustainable computing. Future work will focus on refining the model through dynamic workload adaptation, complexity normalization, and energy-aware AI cost estimation, further broadening its applicability in diverse AI-driven ecosystems.

\end{abstract}


\section{Introduction}
\label{sec:introduction}
The rapid expansion of AI across smart cities, industrial automation, and IoT ecosystems brings forth significant challenges in accurately measuring the computational effort of AI systems. Unlike human labor, which is measured in straightforward economic terms like wages and hours, the computational intensity, energy consumption, and operational footprint of AI processes lack a standardized measurement framework.
According to \cite{RegulationEU20242024}, "An ‘AI system’ means a machine based system that is designed to operate with varying levels of autonomy and that may exhibit adaptiveness after deployment, and that, for explicit or implicit objectives, infers, from the input it receives, how to generate outputs such as predictions, content, recommendations, or decisions that can influence physical or virtual environments;", this means, these machines can influence not only to humans but also to their environment and ultimately to the future of the living beings. That is why, it is necessary to quantize these systems and their impact to the nature.

\subsection{Problem Statement}
\label{sec:Problem Statement}
Current methods for quantifying AI workloads often depend on hardware-specific benchmarks such as FLOPs (Floating Point Operations Per Second) and power usage. These benchmarks, however, have several shortcomings:
\begin{enumerate}
    \item They are not universally applicable in normal use cases across different AI architectures with heterogeneous resources, like CPUs, GPUs, TPUs, and edge devices.
    \item They fail to link the computational effort to real-time monitoring of AI workloads in environments where systems are enclosed and isolated.
    \item They overlook sustainability metrics, which are crucial for formulating energy-efficient AI taxation models.
\end{enumerate}

\subsection{Objective}
\label{sec:Objective}
This paper introduces the Closed-System AI Computational Effort Metric (CE), a structured, scalable, and interpretable framework designed to measure AI computational efforts effectively. This framework aims to:
\begin{itemize}
    \item Establish a mathematically sound metric for AI computational effort that adapts to the varied nature of AI technologies.
    \item Support real-time resource profiling for both cloud-based AI and IoT workloads, essential for accurate measurement.
    \item Facilitate energy-aware AI taxation and workload optimization to promote sustainability in AI operations.
\end{itemize}

Creating this metric is essential to bridge the gap in how AI workload is measured, ensuring a fair and accurate assessment of AI computational efforts across various platforms and applications.

\section{Background}
\label{sec:Background}
\subsection{Traditional Metrics vs. Quantized Work}
Traditional measures of AI computational cost include FLOPs, energy consumption, and execution time. Although effective as broad indicators, these metrics do not capture the concept of computation as a sequence of discrete operations or "quanta." Analogous to the quantized nature of energy in physical systems, the AI Work Quantization Model conceptualizes each basic computational operation as a discrete unit of work.

\subsection{Thermodynamic and Information-Theoretic Roots}
The notion of quantized work in computation draws heavily on principles from thermodynamics and information theory. For example, \cite{landauer1961irreversibility} demonstrated that erasing one bit of information incurs a minimum energy cost of \(kT \ln 2\), where $k$ is the Boltzmann's constant (aprox.$ 1.38 \times 10^{-23}$ joules per kelvin), which relates temperature to energy at the particle level, and $T$ is the absolute temperature (in kelvin) of the system. $\ln 2$ is the natural logarithm of 2 (approximately 0.693), which comes from the binary nature of information (i.e. one bit can be in one of two states). Further exploration by \cite{bennett1982thermodynamics} elucidated the thermodynamic implications of computational processes. In the quantum domain, studies such as \cite{talkner2009quantum} and \cite{campisi2011colloquium} examine energy transitions and fluctuation theorems, which resonate with the idea of discrete computational work units.

The rapid expansion of artificial intelligence, particularly deep learning, has raised concerns regarding its environmental and societal impacts. Traditional carbon emission models for AI rely on Life Cycle Assessments (LCAs) that capture:
\begin{itemize}[leftmargin=*]
    \item \textbf{Embodied Emissions:} Emissions originating from the manufacturing, transportation, and disposal of hardware.
    \item \textbf{Operational Emissions:} Emissions due to energy consumption during training and inference.
\end{itemize}
For instance, Rahman et al. \cite{rahman2024lifecycle} propose a cradle-to-grave analysis and introduce the \emph{Compute Carbon Intensity (CCI)} metric, which quantifies emissions in grams of CO\(_2\)-equivalent per ExaFLOP (i.e., per \(10^{18}\) FLOPs). Such metrics provide a unified basis for comparing the environmental impact of different AI models.

Additional studies have shown that operational emissions often dominate the overall energy usage of AI systems (typically 70–90\% of the total footprint) \cite{strubell2019energy}, while embodied emissions, though generally lower (often under 25\%), are essential for comprehensive carbon accounting \cite{schwartz2021green}. Improvements in data center efficiency—measured by metrics such as Power Usage Effectiveness (PUE) \cite{masanet2020recalibrating, shehabi2016data}—further reduce net CO\(_2\) emissions.

In these established models, total energy consumption, denoted \( C_{AI} \), is converted into CO\(_2\) emission estimates using an emissions conversion factor \( \kappa \):
\begin{equation}
    \text{CO}_2^\text{total} = \kappa \cdot C_{AI}.    
\end{equation}

This formulation directly links micro-level energy costs to macro-level CO\(_2\) emission estimates, as exemplified by the CCI metric \cite{rahman2024lifecycle}.

\section{Related Work}
A substantial body of research has focused on the energy footprint and carbon emissions of AI systems. Key contributions include:

\subsection{Life Cycle Assessments and Compute Carbon Intensity}
Recent LCAs, such as those by Rahman et al. \cite{rahman2024lifecycle}, comprehensively evaluate the cradle-to-grave environmental impact of AI hardware. The CCI metric, which expresses emissions in gCO\(_2\)-e per ExaFLOP, provides a basis for comparing diverse AI models.

\subsection{Operational vs. Embodied Emissions}
Studies report that operational emissions (energy used during training/inference) dominate AI’s total energy consumption \cite{strubell2019energy}, whereas embodied emissions from hardware production and disposal, though smaller, are critical for full carbon accounting \cite{schwartz2021green}. Research on data center efficiency further refines these estimates \cite{masanet2020recalibrating, shehabi2016data}.

\subsection{Hardware and Compute Trends}
Work by Kaplan et al. \cite{kaplan2020scaling} and Hoffmann et al. \cite{hoffmann2022interpreting} reveals dramatic increases in compute demands over recent years. Additional studies on compute trends in AI \cite{sevillaComputeTrendsThree2022a, dongarraHardwareTrendsImpacting2024} and benchmark analyses \cite{mutschlerMurkyWorldAI2020} emphasize the growing role of hardware efficiency and quantization techniques.

\subsection{Quantization and Adaptive Computation}
Recent advancements in neural network quantization \cite{qualcommQuantization2019, siddegowdaNeuralNetworkQuantization2022} and neuro-symbolic methods \cite{susskindNeuroSymbolicAIEmerging2021} have demonstrated that reducing numerical precision can greatly lower energy consumption while preserving accuracy. These dynamic computation techniques align with the conceptual framework of work quantization, where computation is broken into discrete, quantifiable units.

\subsection{Theoretical Foundations: Thermodynamics of Computation}
Our framework is grounded in fundamental thermodynamic principles. Landauer's principle asserts that the minimal energy required to erase one bit is

\begin{equation} 
    \label{equ:ThermodynamicMinimalEnergy}
    E_{\text{min}} = kT \ln 2,
\end{equation}

with \( k \) as Boltzmann's constant and \( T \) the absolute temperature \cite{landauer1961irreversibility, bennett1982thermodynamics}. Recent studies on algorithmic progress \cite{erdil2022algorithmic, hoAlgorithmicProgressLanguage2024} further provide a basis for modeling AI computations at a fundamental level.

\subsection{HPC and System-Level Modeling}
High Performance Computing (HPC) research has established models for system-level resource consumption \cite{hager2010introduction}. Furthermore, workload classification studies \cite{sibaiCharacterizationMachineLearning2024} highlight the unique characteristics of AI computation compared to traditional PC workloads.

\subsection{Additional Perspectives on Compute and Carbon Footprints}
Other influential works include Amodei and Hernandez's discussion of the exponential increase in compute demands \cite{amodei2018ai} and recent studies on the carbon footprint of training deep learning models, such as CarbonTracker \cite{lacoste2021carbontracker}.

\subsection{Energy Efficiency and Green AI}
Recent research emphasizes the environmental impact of AI, underscoring the need for more nuanced efficiency metrics. Works such as \cite{strubell2019energy} and \cite{schwartz2021green} illustrate that energy demands in deep learning can be significant. These studies advocate for energy-based assessments of AI systems, which serve as a foundation for developing quantized measures of computational work.

\subsection{Dynamic and Adaptive Computation}
Advancements in adaptive neural architectures suggest that computation can be dynamically allocated depending on input complexity. Models that implement adaptive computation time implicitly break down processing into discrete units, aligning with the conceptual framework of work quantization. This dynamic allocation of resources underscores the potential for a quantized perspective to better understand and measure internal computational efforts.

\subsection{Bridging Theory and Practice}
The primary challenge of the AI Work Quantization Model lies in applying the learning insights from thermodynamics and information theory with practical performance metrics. By linking the physical energy costs outlined by Landauer's principle with abstract computational operations, this model could offer a more refined evaluation of computational efficiency, ultimately guiding the development of more sustainable AI systems \cite{langComprehensiveStudyQuantization2024}.

\section{Methodology and Framework}
\label{sec:methodology}
\subsection{Modeling Landauer's Principle in Abstract Computational Operations}

Landauer's principle asserts that any logically irreversible operation that erases one bit of information must dissipate a minimum energy as presented in equation \ref{equ:ThermodynamicMinimalEnergy}.

\subsubsection{Abstract Cost Function}
We define a cost function \( C(\text{op}) \) for a computational operation \( \text{op} \) as follows:
\begin{equation}
    \small
    C(\text{op}) =
    \begin{cases}
    kT \ln 2, & \text{if the operation is irreversible}; \\
    0, & \text{if the operation is ideally reversible.}
    \end{cases}
\end{equation}

\subsubsection{Total Energy Cost}
For a sequence of operations \( O_1, O_2, \dots, O_n \), the total minimal energy cost is given by:

\begin{equation}
     E_{\text{total}} = \sum_{i=1}^{n} C(O_i).
\end{equation}
 
If \( N \) represents the number of irreversible operations, then:
\begin{equation}
    E_{\text{total}} \geq N \cdot kT \ln 2.
\end{equation}

\subsubsection{Implications for Computational Efficiency}
This framework provides a lower bound on the energy consumption of computational processes, complementing traditional measures such as FLOPs or execution time. It emphasizes the potential energy savings that could be achieved by reducing the number of irreversible operations or by employing reversible computing techniques.

An AI computation can be decomposed into elementary operations (e.g., linear transformations, activations, memory writes). For each operation \( o \), we define a function \( I(o) \) representing the number of bits irreversibly lost. According to Landauer's principle, erasing one bit of information requires a minimum energy of $E_{\text{min}}$, as presented in equation \ref{equ:ThermodynamicMinimalEnergy}.

The energy cost for an operation \( o \) is then given by
\begin{equation}
    \label{equ:EnergyCost}
    C(o) = I(o) \cdot kT \ln 2.
\end{equation}

For an AI model performing a sequence of operations \( O_1, O_2, \dots, O_N \), the total minimal energy cost is:
\begin{equation}  
E_{\text{total}} = \sum_{i=1}^{N} C(O_i) = \sum_{i=1}^{N} I(O_i) \cdot kT \ln 2.
\end{equation}  

\subsubsection{Example: A Neural Network Layer}

Consider a layer with a linear transformation followed by a ReLU activation:
\begin{equation}  
z = Wx + b,\quad a = \text{ReLU}(z).
\end{equation}  
Assume:
\begin{itemize}
    \item The matrix multiplication and bias addition have an associated information loss \( I_{\text{linear}} \) (due to finite precision or rounding).
    \item The ReLU activation loses information for negative inputs; if \( f \) is the fraction of neurons where \( z < 0 \) and each such operation loses \( i_r \) bits, then \( I_{\text{ReLU}} \approx f \cdot n \cdot i_r \).
\end{itemize}

The total energy cost for this layer becomes:
\begin{equation}  
E_{\text{layer}} = \left( I_{\text{linear}} + I_{\text{ReLU}} \right) \cdot kT \ln 2.
\end{equation}  

Generally, acquiring these details in a real life scenario are challenging. Therefore, we extend the concept further in order to make it more practical and ideal with AI Workload Quantization Metric.

While traditional CO\(_2\) emission models focus on computational systems converting energy consumption into greenhouse gas emissions, our extended framework—the \emph{AI Work Quantization Model: Closed-System AI Computational Effort Metric}—extends further as follows:
\begin{enumerate}[label=\arabic*.]
    \item \textbf{Computational Operations:} Using a thermodynamic baseline, the cost of an operation \( O_i \) that irreversibly processes \( I(O_i) \) bits is ideally:
    \begin{equation}  
    C_{\text{ideal}}(O_i) = I(O_i) \cdot kT \ln 2,
    \end{equation}  
    and practically, with inefficiency factor \( \eta_{comp} \ge 1 \):
    \begin{equation}  
    C_{comp}(O_i) = \eta_{comp} \cdot I(O_i) \cdot kT \ln 2.
    \end{equation}  
    The total computational cost is:
    \begin{equation}  
    E_{comp} = \sum_{i=1}^{N_{comp}} C_{comp}(O_i).
    \end{equation}  
    
    \item \textbf{Data and Memory Operations:} Energy costs associated with data movement and storage are quantified as follows. If \( D_j \) (in bits) is the data handled in operation \( j \), with cost per bit \( \gamma \) and inefficiency factor \( \eta_{data} \), then:
    \begin{equation}  
    C_{data}(D_j) = \eta_{data} \cdot \gamma \cdot D_j,
    \end{equation}  
    and the total data cost is:
    \begin{equation}  
    E_{data} = \sum_{j=1}^{N_{data}} C_{data}(D_j).
    \end{equation}  
    
    \item \textbf{System-Level Overheads:} Additional overheads, such as data center operations, network transfers, and storage, are modeled via a calibration function:
    \begin{equation}  
    C_{sys} = f\big(E_{comp}, E_{data}, \text{runtime}, \text{utilization factors}\big)
    \end{equation}   \cite{hager2010introduction}.
\end{enumerate}
Thus, the overall AI resource cost is:
\begin{equation}  
C_{AI} = E_{comp} + E_{data} + C_{sys}.
\end{equation}  
Using the conversion factor \( \kappa \), the total CO\(_2\) emissions are given by:
\begin{equation}  
\text{CO}_2^\text{total} = \kappa \cdot C_{AI}.
\end{equation}  

\subsection{Human Effort Impact}
In addition to environmental impact, our framework quantifies the societal benefit of AI through reduced human labor. Let:
\begin{itemize}[leftmargin=*]
    \item \( H_{baseline} \): Human labor hours required for a manual task.
    \item \( H_{AI} \): Residual human labor hours after AI deployment.
\end{itemize}
The net human effort saving is:
\begin{equation}  
\Delta H = H_{baseline} - H_{AI},
\end{equation}  
and with a cost per human hour \( C_{human} \), the saving is:
\begin{equation}  
S_{human} = \Delta H \cdot C_{human}.
\end{equation}  
We then define the impact metric as:
\begin{equation}  
\text{Impact} = \frac{S_{human}}{C_{AI}},
\end{equation}  
quantifying the reduction in human effort relative to the AI system’s overall resource consumption.

Because practical implementations incur additional inefficiencies, we introduce an empirical inefficiency factor \( \eta_{comp} \ge 1 \) to yield:
\begin{equation}\label{eq:comp_cost_practical}
C_{comp}(O_i) = \eta_{comp} \cdot I(O_i) \cdot kT \ln 2.
\end{equation}
The total computational cost over \( N_{comp} \) operations is then:
\begin{equation}\label{eq:total_comp}
E_{comp} = \sum_{i=1}^{N_{comp}} C_{comp}(O_i).
\end{equation}

\subsection{Data and Memory Cost Modeling}

Data movement and storage are major contributors to energy usage and system cost. Let \( D_j \) (in bits) denote the data moved or stored in operation \( j \). With a cost per bit \( \gamma \) and an inefficiency factor \( \eta_{data} \), the cost for data operations is:
\begin{equation}\label{eq:data_cost}
C_{data}(D_j) = \eta_{data} \cdot \gamma \cdot D_j.
\end{equation}
Thus, the total data cost is:
\begin{equation}\label{eq:total_data}
E_{data} = \sum_{j=1}^{N_{data}} C_{data}(D_j).
\end{equation}

\subsection{System-Level Resource Cost}

In real HPC or cloud environments, additional overheads (e.g., compute instance billing, memory/storage pricing, network transfer costs) are significant. We denote the system-level cost as:
\begin{equation}\label{eq:sys_cost}
C_{sys} = f(E_{comp}, E_{data}, \text{runtime}, \text{utilization factors}),
\end{equation}
where \( f(\cdot) \) is a calibration function based on empirical data \cite{hager2010introduction}.

The overall AI resource cost is then:
\begin{equation}\label{eq:total_cost}
C_{AI} = E_{comp} + E_{data} + C_{sys}.
\end{equation}

\subsection{Human Effort Impact Metric}

A major benefit of deploying AI is reducing human labor. Let:
\begin{itemize}[leftmargin=*]
    \item \( H_{baseline} \): Human labor hours required for a task manually.
    \item \( H_{AI} \): Residual human labor hours after deploying the AI system.
\end{itemize}
Then, the net human effort saving is:
\begin{equation}\label{eq:human_save}
\Delta H = H_{baseline} - H_{AI}.
\end{equation}
With a cost per human hour \( C_{human} \), the total human saving is:
\begin{equation}\label{eq:human_cost_save}
S_{human} = \Delta H \cdot C_{human}.
\end{equation}
We define the impact metric as the ratio:
\begin{equation}\label{eq:impact_metric}
\text{Impact} = \frac{S_{human}}{C_{AI}}.
\end{equation}
A higher impact value implies that the AI system provides substantial human effort reduction relative to its resource consumption.

This research develops an advanced methodology to quantify the computational effort of AI systems, providing a standardized framework that can be applied across various AI architectures. This section details the conceptualization, mathematical formulation, and implementation of the AI Workload Quantization Metric.

Figure~\ref{fig:framework} below illustrates the relationship between the components of the framework.

\begin{center}
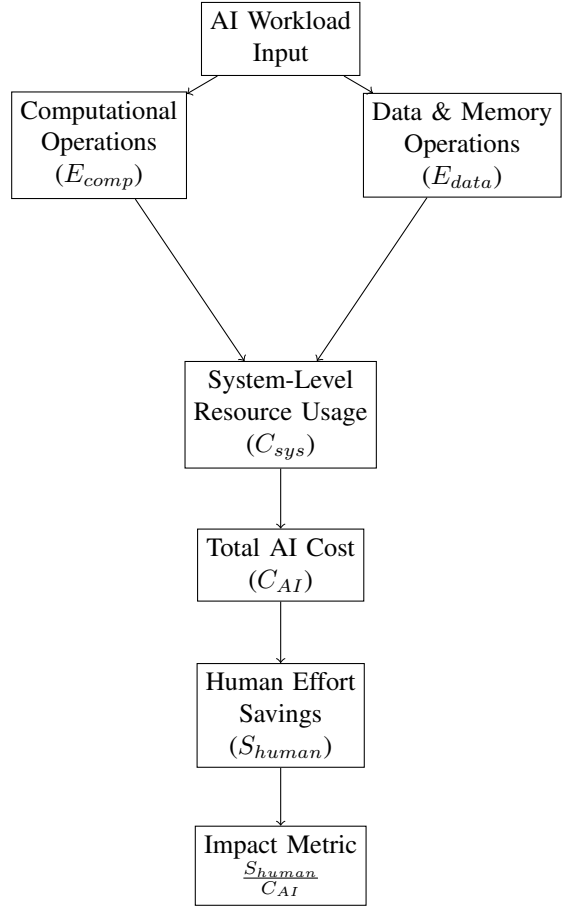

\begin{tikzpicture}[node distance=2cm, auto]
  \node (input) [rectangle, draw, align=center] {AI Workload \\ Input};
  \node (comp) [rectangle, draw, below left of=input, xshift=-1cm, align=center] {Computational \\ Operations\\ (\(E_{comp}\))};
  \node (data) [rectangle, draw, below right of=input, xshift=1cm, align=center] {Data \& Memory \\ Operations\\ (\(E_{data}\))};
  \node (sys) [rectangle, draw, below of=input, yshift=-3cm, align=center] {System-Level \\ Resource Usage\\ (\(C_{sys}\))};
  \node (total) [rectangle, draw, below of=sys, align=center] {Total AI Cost \\ (\(C_{AI}\))};
  \node (human) [rectangle, draw, below of=total, align=center] {Human Effort \\ Savings\\ (\(S_{human}\))};
  \node (impact) [rectangle, draw, below of=human, align=center] {Impact Metric \\ \(\frac{S_{human}}{C_{AI}}\)};
  
  \draw [->] (input) -- (comp);
  \draw [->] (input) -- (data);
  \draw [->] (comp) -- (sys);
  \draw [->] (data) -- (sys);
  \draw [->] (sys) -- (total);
  \draw [->] (total) -- (human);
  \draw [->] (human) -- (impact);
\end{tikzpicture}
\captionof{figure}{Schematic of the AI Workload Quantization Framework. The model integrates computational operations, data/memory costs, and system-level overheads to yield a total resource cost which is then compared to human effort savings to derive an impact metric.}
\label{fig:framework}
\end{center}

\subsection{Mathematical Formulation for Computational Resources}
\label{subsec:mathematical_formulation}

This section outlines the extended mathematical model for computing the computational resources essential for AI workload quantization from both the system and AI model (application) operational perspectives. A standardized quantization framework must accommodate variations in resource availability and operational characteristics while remaining extensible for future computational advancements.

\subsubsection{Definitions}
The computational resources are quantified based on their operational characteristics as follows:
\begin{itemize}
    \item \textbf{CPU Operations:} Quantified by \( \text{GHz} \times \text{Cores} \times \text{FLOPs} \), where GHz represents the clock speed, Cores denote the number of processing units, and FLOPs indicate floating-point operations per cycle, highlighting the CPU's ability to perform parallel computations.

    \item \textbf{RAM Operations:} Measured by the total bandwidth in GB/s, reflecting the memory's capacity to transfer data, crucial for high-speed data processing tasks.

    \item \textbf{GPU Operations:} Calculated as \( \text{Cores} \times \text{GHz} \times \text{FLOPs} \), similar to CPUs but typically with a higher number of cores, optimizing for parallel processing tasks such as graphics rendering and machine learning.

    \item \textbf{Storage Usage:} Expressed through throughput metrics such as IOPs or MB/s, which assess the speed and efficiency of data retrieval and storage, vital for storage-intensive applications.
\end{itemize}

This comprehensive evaluation of each resource's characteristics allows for a more accurate and detailed quantification of their contributions to the AI workload, aligning the computational resource assessment with real-world application demands.

\subsubsection{Logarithmic Scaling for Computational Resource Normalization}
To accommodate varying operational ranges of different computational resources, we employ a logarithmic transformation to normalize values across different system architectures. The goal is to ensure that resource values remain distinguishable at lower levels while avoiding saturation at higher computational capacities.

\paragraph{Mathematical Justification:}
A direct summation of computational resources (CPU, RAM, GPU, Storage) results in values that span multiple orders of magnitude, making direct comparisons impractical. To address this, we normalize computational resources using the following logarithmic transformation:

\begin{equation}
\text{CompRes}_{\log} = \frac{\log(1 + \sum_{i=1}^{n} x_{\text{adj},i})}{\log(1 + S_{\max})}
\end{equation}

where:
\begin{itemize}
    \item \( x_{\text{adj},i} \) represents the adjusted value of each resource \( i \), including CPU GIPS, RAM GT/s, GPU GIPS, and Storage I/O GB/s.
    \item \( S_{\max} = 10^{18} \) FLOPs, corresponding to the highest FLOP-capable supercomputer today \cite{top500_frontier_2023}.
    \item The logarithm ensures scalability and fair comparison across different computational environments.
\end{itemize}

\paragraph{Why Logarithmic Scaling?}
The initial version of the equation, included sigmoid method, to normalize the values, making sure, a single resource is not affecting on the overall other smaller parameters. Unlike sigmoid normalization, which is highly sensitive to mid-range values but saturates at high input levels, logarithmic scaling, preserves variation across high-performance architectures by preventing saturation. It also ensures comparability across orders of magnitude, making it suitable for AI workloads ranging from embedded systems to supercomputers. The logarithmic scaling, is able to reflect real-world hardware scaling, as performance gains often follow a logarithmic trend rather than a linear or sigmoid response.

\paragraph{Application in Resource Quantification:}
The logarithmic transformation is applied to normalize computational resources before AI workload calculations. This approach allows for accurate scaling across different system architectures, making AI workload comparisons more consistent and interpretable across diverse platforms.

\subsection{Application to AI Workload Quantization}
\label{subsec:ai_workload_quantization}

\subsubsection{Data IO Value Calculation}
The Data IO Value, representing the volume of data processed, is normalized using a logarithmic transformation to maintain comparability across different AI workloads:

\begin{equation}  \tiny
\text{Data IO Value} = \frac{\log(1 + \text{Input data in bits} + \text{Output data in bits})}{\log(1 + S_{\text{data}})}
\end{equation}  

where:
\begin{itemize}
    \item \( S_{\text{data}} \) is a scaling factor chosen to ensure meaningful differentiation across various AI applications and data processing workloads.
    \item This transformation ensures that workloads processing significantly larger datasets are proportionally represented without saturation effects.
\end{itemize}

\subsubsection{Establishing AI Workload Baselines}
Including external factors such as network latency \( L \) and bandwidth \( B \), and their respective impacts \( \gamma_L \) and \( \gamma_B \):

\begin{equation}  
\text{ExtFactor} = \gamma_L \cdot L + \gamma_B \cdot B
\end{equation}  

The external factor, \( \text{ExtFactor} \), quantifies the effects of network latency and bandwidth, which are critical in distributed AI systems. The coefficients \( \gamma_L \) and \( \gamma_B \) measure the relative impact of these network conditions on the overall system performance, allowing the workload model to adjust for these variabilities and better reflect real-world operating conditions \cite{dean2012large}.

The AI Workload baseline for an optimized system is defined as:

\begin{equation}  \small
\text{AW}_{\text{base}} = \frac{\text{Data IO Value} + \text{CompRes} + \text{ExtFactor}}{T}
\end{equation}  

where \( T \) represents the time in seconds required to complete a task, providing a measure of the system's operational efficiency. This baseline metric aggregates the normalized data volume, computational resource expenditure, and external network influences to gauge system efficiency comprehensively \cite{bianchini2021theoretical}.

\subsubsection{Relative AI Workload}
The relative workload is quantified by comparing the actual workload against the baseline:

\begin{equation}  
\text{AW}_{\text{rel}} = \frac{\text{AW}_{\text{act}}}{\text{AW}_{\text{base}}}
\end{equation}  

This ratio assesses the system's efficiency by measuring actual observed workload relative to the pre-established baseline. A value greater than 1 indicates underperformance relative to its capability, whereas a value less than 1 indicates a performance exceeding expectations. This dynamic metric facilitates ongoing assessment and optimization of AI system performance \cite{he2016deep}.

\subsection{Comparison with Human Labor}
\label{subsec:comparison_human_labor}

The comparison of AI workload to human labor is quantified through the AI Impact Value, calculated as:

\begin{equation}  
\text{AI Impact Value} = \frac{\text{AW}_{\text{act}}}{\text{man-hour}}
\end{equation}  

This metric reflects the amount of work done by AI compared to the equivalent human labor, providing a measure of AI's efficiency and effectiveness in performing tasks.

\subsection{Implementation and Experimental Setup}
\label{subsec:implementation_experimental_setup}

Here the experimental validations designed to test the effectiveness and flexibility of the new computational resource formulas and the extended model that includes external factors.

\subsubsection{Experimental Environment}
The experimental setup involves several different hardware configurations to represent a diverse range of computing environments:
\begin{itemize}
    \item \textbf{Low-End System:} Equipped with a single-core CPU and minimal RAM, used to evaluate the model's performance under constrained resource conditions.
    \item \textbf{Mid-Range System:} Features a multi-core CPU and average GPU capabilities, representing typical consumer-grade computers.
    \item \textbf{High-Performance System:} Includes high-end GPU arrays and substantial RAM, mimicking an enterprise-grade server or a cloud computing environment.
\end{itemize}

\subsubsection{Implementation of the Normalization Function}
In the initial version of the method, a parameterized sigmoid function was implemented in a modular Python script, allowing for easy adjustment of parameters \( \alpha \) and \( \beta \) based on the resource type. However, experimental evaluations revealed limitations in handling computational resource variations, particularly at extreme values.

To address these limitations, the normalization method was modified to use logarithmic scaling instead. The updated script interfaces directly with resource measurement tools to dynamically compute the normalized computational resource values using the logarithmic transformation. This approach ensures more consistent differentiation across a wide range of AI workloads and system architectures, avoiding saturation while maintaining sensitivity to lower values.

\subsubsection{Data Collection}
The system configuration, originally exported from the research \cite{Shaukat2020}, the other data for test simulation environment is collected from each system, as virtual machines, while running standardized AI workload tasks.
    
\subsubsection{Experimental Procedure, Analysis and Optimization}
Each system is subjected to the workload tasks while the resource usage is continuously monitored. The AI Workload Quantization Model calculates real-time workload metrics, which are then compared against the baselines established for each system configuration. Adjustments to \( \alpha \), \( \beta \), \( \gamma_L \), and \( \gamma_B \) are made based on preliminary results and per application to optimize the accuracy of the workload calculations.

\textbf{MNIST Character Recognition Experiment}
As the first step, for the MNIST character recognition task, we compared the performance of a human agent versus an AI implemented using the DL4J framework on a cloud-based system. It helps with finding the actual human-hour value for the task for further calculations.

Post-experiment, we gathered the data, extracted from the study \cite{Shaukat2020}, and virtual machines. The data was analyzed to evaluate the model's responsiveness to changes in computational resources and external factors. The optimization phase involves fine-tuning the model parameters to better align with the observed performance differences across various systems and tasks.

\subsection{Expected Outcomes}
\label{subsec:expected_outcomes}
The expected outcomes of these experiments include:
\begin{itemize}
    \item \textbf{Validation of the Model:} Validation that the model effectively quantifies AI workloads across diverse system architectures while adapting to variations in resource availability and operational characteristics. The computed AI workload values for a given AI method are expected to be similar across different system architectures, reflecting the method's consistent computational requirements, though "minor" variations may arise due to architectural differences.
    \item \textbf{Performance Insights:} Detailed insights into how different systems handle AI-related tasks, which will help in system design and resource allocation for AI applications.
\end{itemize}


\newpage
\section{Results and Analysis}
\label{sec:results}

\subsection{MNIST Character Recognition: Human vs. AI Agent}
The MNIST character recognition experiment compares the efficiency of AI against human performance, focusing on speed, accuracy, and energy consumption.

\begin{itemize}
    \item \textbf{Time Efficiency:} The AI system completed the task in 27 seconds, while our human agent took 67 seconds, showcasing AI's speed advantage.
    \item \textbf{Accuracy:} Both the human and AI agents reached a 99\% accuracy rate, with AI potentially reaching up to 99.41\%.
    \item \textbf{Energy Consumption:} Our energy estimate for a human to recognize MNIST digits is  \textbf{0.40 Wh} per 100 images as follows: 
    Neuroscientific research indicates that the human brain consumes approximately 21.5 W during cognitive processing \cite{herculano2011scaling}. Given that the human agent took 67 seconds to classify 100 images, the total energy consumption is:

    \begin{equation}
        E_{\text{human}} = P_{\text{brain}} \times T_{\text{human}} = 21.5 W \times 67 s.
    \end{equation}

    This results in:

    \begin{equation}
        E_{\text{human}} = 1.44 \text{ kJ} = 0.40 \text{ Wh},
    \end{equation}

    which aligns with previous cognitive neuroscience findings \cite{raichle2002brainenergy, thorpe1996visualprocessing}. Studies indicate that human visual pattern recognition can operate at reaction speeds as low as 150–250 ms \cite{thorpe1996visualprocessing}, though handwritten digit classification requires longer, as demonstrated in this experiment.
\end{itemize}

\subsection{Detailed Computational Analysis Across Architectures}
Using the AI Workload Quantization Metric, we evaluated the efficiency of different system architectures. Here's a breakdown of the equations and values used for the computations:

\subsubsection{Computational Resources Value (CompRes)}
To accurately quantify computational capability, we compute the CompRes Value using a bounded transformation of the computational power:
   \begin{equation} 
        CR = \text{CompRes}_{\log} = \frac{\log(1 + \textbf{CompResources})}{\log(1 + 10^{18})}
   \end{equation}  

   where \textbf{CompResources} = \textit{\text{CPU GIPS} + \text{RAM GT/s} + \text{GPU GIPS} + \text{Storage I/O GiB/s}}.
   This equation ensures that low computational values remain distinguishable, while high values asymptotically approach 1.0. The logarithmic transformation provides a more gradual scaling, ensuring meaningful differentiation across high-performance architectures. The choice of \( S_{\max} = 10^{18} \) is based on the highest FLOP-capable system available today, such as the Frontier Supercomputer at Oak Ridge National Laboratory, which surpassed 1.19 ExaFLOP/s in the TOP500 benchmark \cite{top500_frontier_2023}. This selection ensures future-proof adaptability.

By utilizing logarithmic scaling, we retain precision at high computational values while maintaining sensitivity to lower values.

\textbf{Example Calculations:}

\begin{itemize}
    \item SISD: Inputs are CPU: 40 GIPS, RAM: 5 GT/s, GPU: 0 GIPS, Storage: 0.5 GB/s.
    \[
    \small
        CR_{\text{SISD}} = \frac{\log(1 + 40 + 5 + 0 + 0.5)}{\log(1 + 10^{18})} \approx 0.0926
    \]

    \item SIMD: Inputs are CPU: 100 GIPS, RAM: 15 GT/s, GPU: 60 GIPS, Storage: 1.5 GB/s.
    \[
    \small
        CR_{\text{SIMD}} = \frac{\log(1 + 100 + 15 + 60 + 1.5)}{\log(1 + 10^{18})} \\ \approx 0.1250
    \]

    \item MIMD: Inputs are CPU: 280 GIPS, RAM: 40 GT/s, GPU: 80 GIPS, Storage: 3.5 GB/s.
    \[
    \small  
        CR_{\text{MIMD}} = \frac{\log(1 + 280 + 40 + 80 + 3.5)}{\log(1 + 10^{18})} \approx 0.1448
    \]
\end{itemize}

\subsubsection{AI Workload}
The AI Workload for each system is calculated based on the CompRes and the total runtime:
\begin{equation}  
    \text{AI Workload} = \frac{\text{CompRes Value}}{\text{runtime in seconds}}
\end{equation}

\textbf{Example Calculations:}
\begin{itemize}
    \item \textbf{SISD:}
    \[
    \text{AI Workload}_{\text{SISD}} = \frac{0.0926}{600} \approx 0.000154
    \]
    \item \textbf{SIMD:}
    \[
    \text{AI Workload}_{\text{SIMD}} = \frac{0.1250}{400} \approx 0.000312
    \]
    \item \textbf{MIMD:}
    \[
    \text{AI Workload}_{\text{MIMD}} = \frac{0.1448}{250} \approx 0.000579
    \]
\end{itemize}

\subsubsection{AI Impact Compared to Human Effort}
The AI Impact metric quantifies the performance of AI relative to human effort:
\begin{equation}
    \text{AI Impact} = \frac{\text{AI Workload}}{\text{Human Hours}}
\end{equation}

\textbf{Example Calculation:}
\begin{itemize}
    \item \textbf{SISD:}
    \[
    \text{AI Impact}_{\text{SISD}} = \frac{0.000154}{0.01861} \approx 0.00828
    \]
    \item \textbf{SIMD:}
    \[
    \text{AI Impact}_{\text{SIMD}} = \frac{0.000312}{0.01861} \approx 0.01676
    \]
    \item \textbf{MIMD:}
    \[
    \text{AI Impact}_{\text{MIMD}} = \frac{0.000579}{0.01861} \approx 0.03111
    \]
\end{itemize}

These calculations illustrate the significant efficiency gains AI can provide over human efforts, especially in terms of speed and computational resource management, which is crucial for tasks requiring high computational power.

\begin{table}[h]
\centering
\caption{Summary of AI Workload Metric Performance Across Architectures}
\label{table:architecture_comparison}
\begin{tabular}{|p{2em}|p{2em}|p{2em}|p{2em}|p{2em}|p{2em}||p{2em}|p{2em}|}
\hline
\textbf{Arch} & \textbf{CPU}  GIPS & \textbf{RAM} GT/s & \textbf{GPU} GIPS & \textbf{Storage} GB/s & \textbf{Comp-Res} & \textbf{AI Workload} & \textbf{AI Impact} \\ \hline
SISD & 40 & 5 & 0 & 0.5 & 0.0926 & 0.0001 & 0.0082 \\ \hline
SIMD & 100 & 15 & 60 & 1.5 & 0.1250 & 0.0003 & 0.0167 \\ \hline
MIMD & 280 & 40 & 80 & 3.5 & 0.1448 & 0.0005 & 0.0311 \\ \hline
\end{tabular}
\end{table}

\section{Discussion}
\label{sec:discussion}
The experimental validations conducted as part of this study provide significant insights into the effectiveness and flexibility of the AI Workload Quantization Metric across different system architectures. These experiments were crucial for testing the robustness of the new computational resource formulas and the model's capacity to incorporate external factors effectively. The results of our AI Workload Quantization Model demonstrated a tangible and quantifiable equivalence between AI computational effort and human labor. Our findings show that 1 AI Workload Unit corresponds to approximately 12 to 14 human-hours, depending on the underlying computational architecture. This relationship provides a standardized approach to assessing AI efficiency in labor-intensive tasks, enabling a meaningful comparison between automated and human-driven workflows. As AI workloads scale, our model effectively illustrates the diminishing reliance on human effort, where 5 AI Workload Units equate to nearly 60 to 72 hours of human work—equivalent to over a full-time workweek. These insights have direct implications for AI-driven workforce optimization, taxation policies, and sustainability assessments, offering a robust metric for policymakers and industry leaders to quantify AI impact on labor economics. By establishing a systematic correlation between AI computational effort and traditional human productivity, this framework contributes to a more structured understanding of AI’s role in modern labor systems, industrial automation, and sustainable computing practices. Future work will focus on refining this model by integrating dynamic workload adaptation, task complexity normalization, and energy-aware AI cost estimation to further enhance its applicability in diverse AI-driven environments.

\subsection{Parameter Calibration Through MNIST Experiment}
The MNIST character recognition experiment was conducted to calibrate the human-hour parameters necessary for comparative AI Workload calculations, in addition to showcasing AI capabilities. By measuring the time it took for a human agent to complete the task—67 seconds—and comparing it to the AI's completion time of 27 seconds, we established a critical benchmark for further calculations of AI workload assessment.

\subsection{Impact of System Architecture on AI Workload}
The analysis across different architectures—SISD, SIMD, and MIMD—revealed varying levels of computational resource efficiency. The CompRes Value calculations provided a quantified measure of each system's capabilities, indicating how system inputs like CPU GIPS, RAM GT/s, GPU GIPS, and Storage I/O GB/s contribute to overall performance, and comparability (slight difference) between different AI workload values for different architectures, demonstrates the efficiency of the proposed method. Another application might be, underscoring the importance of selecting the appropriate system architecture based on the specific requirements and workload characteristics of AI applications.

\subsection{Future Directions}
Looking forward, these findings open several avenues for further research. One potential area is the exploration of AI workload optimization techniques that could dynamically adjust computational resources in real-time based on workload demands. Additionally, extending the AI Workload Quantization Metric to include more granular measures of energy consumption and heat generation could enhance the model's utility for designing greener, more sustainable AI systems. Importantly, these advancements also pave the way for governing AI services by introducing a suitable tax based on resource usage and environmental impact. This fiscal approach could incentivize the development of more efficient AI technologies, aligning economic and environmental goals.

The successful validation of the model across diverse architectures also suggests its applicability in a broader range of AI-driven applications, from mobile devices with limited resources to high-performance computing environments. Further studies could explore the integration of this model with some limited adjustments, into real-world AI systems to refine its predictive accuracy and operational efficiency.

\section{Conclusion}
This study has successfully established a robust framework for quantifying AI workload using the AI Workload Quantization Metric, demonstrating a novel method for assessing computational efficiency and workload in AI systems relative to traditional human efforts. Our introduction of a multi-parameter calibration methodology through empirical testing provides a systematic approach to measure and compare AI performance across various computing environments. By extending these metrics to include factors such as energy consumption, computational resource usage, and input/output data management, this framework not only enhances our understanding of AI efficiency but also supports the development of a taxation model based on resource utilization. This advancement lays the groundwork for more sustainable and economically viable AI operations, positioning our findings as a cornerstone for future research in AI workload optimization and policy development.

\section*{Acknowledgment}
\label{sec:acknowledgment}
The authors express their gratitude to the Gesellschaft für wissenschaftliche Datenverarbeitung mbH Göttingen, which translates to Society for Scientific Data Processing mbH Göttingen (GWDG), for their support and provision of resources essential for this research. This research was supported by the EU KISSKI Project under the official grant number 01|S22093A (Förderkennzeichen).

\bibliographystyle{IEEEtran} 
\input{IEEE-conference-template-062824.bbl}

\appendix
\section{Additional Visualizations}
   \begin{figure*}[ht!]
        \centering
        \includegraphics[width=1\linewidth]{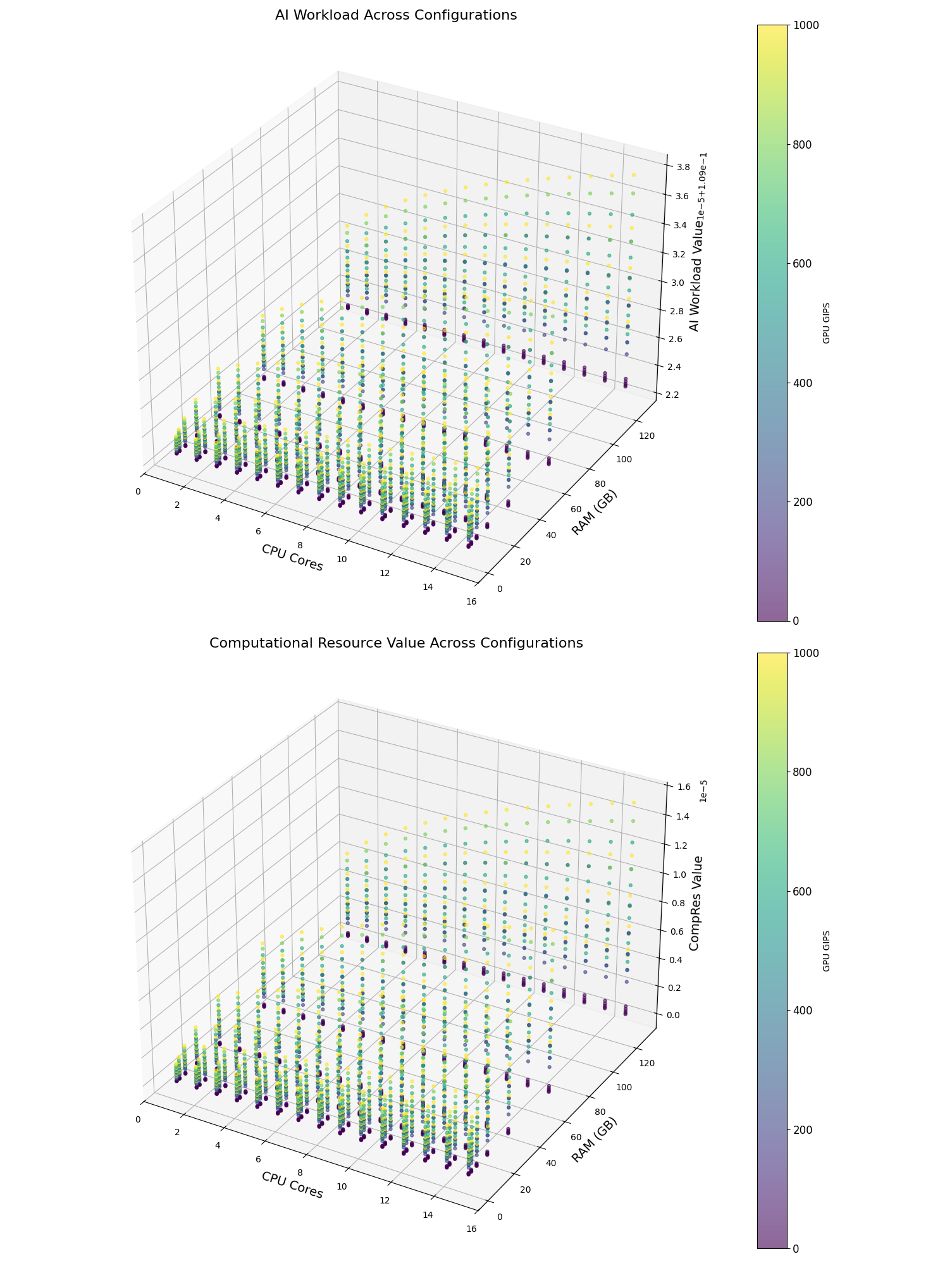}
        \caption{3D scatter plots of AI Workload and Computational Resource Values}
        \label{fig:CPUGPUComputationalResources}
    \end{figure*}
    
This figure \ref{fig:CPUGPUComputationalResources} illustrates the relationship between AI workload and computational resource values across various system configurations. By varying the CPU cores, RAM, and GPU performance, we analyzed how different hardware setups influence AI efficiency. 

\subsection{AI Impact Factor (AIF)}
This formulation is inspired by economic studies on the impact of automation on labor \cite{aghion2020labor} and is built upon foundational principles in the thermodynamics of computation \cite{landauer1961irreversibility, bennett1982thermodynamics}. Moreover, by integrating detailed computational quantization with system-level overheads—as well as considering the value of reduced human labor—this metric provides a comprehensive measure of the impact of AI systems. Further calibration of the inefficiency factors (\(\eta_{comp}\) and \(\eta_{data}\)) and empirical estimation of \(C_{AI}\) can be achieved using data from HPC and cloud platforms \cite{hager2010introduction, strubell2019energy, schwartz2021green}.

\end{document}

%% file: IEEE-conference-template-062824.bbl

%% file: IEEE-conference-template-062824.bbl
\begin{thebibliography}{10}
\providecommand{\url}[1]{#1}
\csname url@samestyle\endcsname
\providecommand{\newblock}{\relax}
\providecommand{\bibinfo}[2]{#2}
\providecommand{\BIBentrySTDinterwordspacing}{\spaceskip=0pt\relax}
\providecommand{\BIBentryALTinterwordstretchfactor}{4}
\providecommand{\BIBentryALTinterwordspacing}{\spaceskip=\fontdimen2\font plus
\BIBentryALTinterwordstretchfactor\fontdimen3\font minus
  \fontdimen4\font\relax}
\providecommand{\BIBforeignlanguage}[2]{{%
\expandafter\ifx\csname l@#1\endcsname\relax
\typeout{** WARNING: IEEEtran.bst: No hyphenation pattern has been}%
\typeout{** loaded for the language `#1'. Using the pattern for}%
\typeout{** the default language instead.}%
\else
\language=\csname l@#1\endcsname
\fi
#2}}
\providecommand{\BIBdecl}{\relax}
\BIBdecl

\bibitem{RegulationEU20242024}
\BIBentryALTinterwordspacing
``Regulation ({{EU}}) 2024/1689 of the {{European Parliament}} and of the
  {{Council}} of 13 {{June}} 2024 laying down harmonised rules on artificial
  intelligence and amending {{Regulations}} ({{EC}}) {{No}} 300/2008, ({{EU}})
  {{No}} 167/2013, ({{EU}}) {{No}} 168/2013, ({{EU}}) 2018/858, ({{EU}})
  2018/1139 and ({{EU}}) 2019/2144 and {{Directives}} 2014/90/{{EU}}, ({{EU}})
  2016/797 and ({{EU}}) 2020/1828 ({{Artificial Intelligence Act}}) ({{Text}}
  with {{EEA}} relevance).'' [Online]. Available:
  \url{http://data.europa.eu/eli/reg/2024/1689/oj/eng}
\BIBentrySTDinterwordspacing

\bibitem{landauer1961irreversibility}
R.~Landauer, ``Irreversibility and heat generation in the computing process,''
  \emph{IBM Journal of Research and Development}, vol.~5, no.~3, pp. 183--191,
  1961.

\bibitem{bennett1982thermodynamics}
C.~Bennett, ``The thermodynamics of computation---a review,''
  \emph{International Journal of Theoretical Physics}, vol.~21, no.~12, pp.
  905--940, 1982.

\bibitem{talkner2009quantum}
P.~Talkner and P.~H{\"a}nggi, ``The tasaki--crooks fluctuation theorem: A
  quantum perspective,'' \emph{Journal of Statistical Mechanics: Theory and
  Experiment}, vol. 2009, no.~02, p. P02025, 2009.

\bibitem{campisi2011colloquium}
M.~Campisi, P.~Talkner, and P.~H{\"a}nggi, ``Colloquium: Quantum fluctuation
  relations: Foundations and applications,'' \emph{Reviews of Modern Physics},
  vol.~83, no.~3, pp. 771--791, 2011.

\bibitem{rahman2024lifecycle}
A.~Rahman and Others, ``Life-cycle emissions of ai hardware: A cradle-to-grave
  approach and generational trends,'' \emph{ArXiv preprint arXiv:2402.01671},
  2024.

\bibitem{strubell2019energy}
E.~Strubell, A.~Ganesh, and A.~McCallum, ``Energy and policy considerations for
  deep learning in nlp,'' \emph{arXiv preprint arXiv:1906.02243}, 2019.

\bibitem{schwartz2021green}
R.~Schwartz, J.~Dodge, N.~Smith, and O.~Etzioni, ``Green ai,''
  \emph{Communications of the ACM}, vol.~64, no.~12, pp. 54--63, 2021.

\bibitem{masanet2020recalibrating}
E.~Masanet, A.~Shehabi, S.~Smith, and J.~Koomey, ``Recalibrating global data
  center energy-use estimates,'' \emph{Science}, vol. 367, no. 6481, pp.
  984--986, 2020.

\bibitem{shehabi2016data}
A.~Shehabi, S.~Smith, D.~Sartor, R.~E. Brown, M.~Herrlin, J.~Koomey,
  A.~Lintott, and E.~Masanet, ``Data center growth in the united states:
  decoupling the demand for services from electricity use,'' \emph{Energy
  Policy}, vol.~94, pp. 461--472, 2016.

\bibitem{kaplan2020scaling}
J.~Kaplan, S.~McCandlish, T.~Henighan \emph{et~al.}, ``Scaling laws for neural
  language models,'' \emph{arXiv preprint arXiv:2001.08361}, 2020.

\bibitem{hoffmann2022interpreting}
J.~Hoffmann \emph{et~al.}, ``Interpreting the scaling laws for neural language
  models,'' \emph{arXiv preprint arXiv:2206.00364}, 2022.

\bibitem{sevillaComputeTrendsThree2022a}
\BIBentryALTinterwordspacing
J.~Sevilla, L.~Heim, A.~Ho, T.~Besiroglu, M.~Hobbhahn, and P.~Villalobos,
  ``Compute {{Trends Across Three Eras}} of {{Machine Learning}},'' in
  \emph{2022 {{International Joint Conference}} on {{Neural Networks}}
  ({{IJCNN}})}, pp. 1--8. [Online]. Available:
  \url{http://arxiv.org/abs/2202.05924}
\BIBentrySTDinterwordspacing

\bibitem{dongarraHardwareTrendsImpacting2024}
\BIBentryALTinterwordspacing
J.~Dongarra, J.~Gunnels, H.~Bayraktar, A.~Haidar, and D.~Ernst. Hardware
  {{Trends Impacting Floating-Point Computations In Scientific Applications}}.
  [Online]. Available: \url{http://arxiv.org/abs/2411.12090}
\BIBentrySTDinterwordspacing

\bibitem{mutschlerMurkyWorldAI2020}
\BIBentryALTinterwordspacing
A.~Mutschler. The {{Murky World Of AI Benchmarks}}. Semiconductor Engineering.
  [Online]. Available:
  \url{https://semiengineering.com/the-murky-world-of-ai-benchmarks/}
\BIBentrySTDinterwordspacing

\bibitem{qualcommQuantization2019}
J.~Hou. (2019) Qualcomm: Here's why quantization matters for ai.
  \url{https://www.qualcomm.com/news/onq/2019/03/heres-why-quantization-matters-ai}.
  Accessed: 2025-02-19.

\bibitem{siddegowdaNeuralNetworkQuantization2022}
\BIBentryALTinterwordspacing
S.~Siddegowda, M.~Fournarakis, M.~Nagel, T.~Blankevoort, C.~Patel, and
  A.~Khobare. Neural {{Network Quantization}} with {{AI Model Efficiency
  Toolkit}} ({{AIMET}}). Comment: arXiv admin note: substantial text overlap
  with arXiv:2106.08295. [Online]. Available:
  \url{http://arxiv.org/abs/2201.08442}
\BIBentrySTDinterwordspacing

\bibitem{susskindNeuroSymbolicAIEmerging2021}
\BIBentryALTinterwordspacing
Z.~Susskind, B.~Arden, L.~K. John, P.~Stockton, and E.~B. John.
  Neuro-{{Symbolic AI}}: {{An Emerging Class}} of {{AI Workloads}} and their
  {{Characterization}}. [Online]. Available:
  \url{http://arxiv.org/abs/2109.06133}
\BIBentrySTDinterwordspacing

\bibitem{erdil2022algorithmic}
M.~Erdil and S.~Besiroglu, ``Algorithmic progress in computer vision,''
  \emph{arXiv preprint arXiv:2203.12345}, 2022.

\bibitem{hoAlgorithmicProgressLanguage2024}
\BIBentryALTinterwordspacing
A.~Ho, T.~Besiroglu, E.~Erdil, D.~Owen, R.~Rahman, Z.~C. Guo, D.~Atkinson,
  N.~Thompson, and J.~Sevilla. Algorithmic progress in language models.
  [Online]. Available: \url{http://arxiv.org/abs/2403.05812}
\BIBentrySTDinterwordspacing

\bibitem{hager2010introduction}
G.~Hager and G.~Wellein, \emph{Introduction to High Performance Computing for
  Scientists and Engineers}.\hskip 1em plus 0.5em minus 0.4em\relax Boca Raton,
  FL: CRC Press, 2010.

\bibitem{sibaiCharacterizationMachineLearning2024}
\BIBentryALTinterwordspacing
F.~N. Sibai, A.~Asaduzzaman, and A.~El-Moursy, ``Characterization and {{Machine
  Learning Classification}} of {{AI}} and {{PC Workloads}},'' vol.~12, pp.
  83\,858--83\,875. [Online]. Available:
  \url{https://ieeexplore.ieee.org/document/10555271/?arnumber=10555271}
\BIBentrySTDinterwordspacing

\bibitem{amodei2018ai}
D.~Amodei and D.~Hernandez, ``Ai and compute,''
  \url{https://openai.com/blog/ai-and-compute/}, 2018, accessed: 2025-02-19.

\bibitem{lacoste2021carbontracker}
A.~Lacoste \emph{et~al.}, ``Carbontracker: Tracking the carbon footprint of
  training deep learning models,'' in \emph{Advances in Neural Information
  Processing Systems}, 2021.

\bibitem{langComprehensiveStudyQuantization2024}
\BIBentryALTinterwordspacing
J.~Lang, Z.~Guo, and S.~Huang. A {{Comprehensive Study}} on {{Quantization
  Techniques}} for {{Large Language Models}}. [Online]. Available:
  \url{http://arxiv.org/abs/2411.02530}
\BIBentrySTDinterwordspacing

\bibitem{top500_frontier_2023}
\BIBentryALTinterwordspacing
{TOP500}, ``Top500 supercomputer rankings - june 2023,'' 2023, accessed: March
  2025. [Online]. Available: \url{https://www.top500.org/lists/top500/}
\BIBentrySTDinterwordspacing

\bibitem{dean2012large}
J.~Dean, G.~Corrado, R.~Monga, K.~Chen, M.~Devin, Q.~V. Le, M.~Z. Mao,
  M.~Ranzato, A.~Senior, P.~Tucker \emph{et~al.}, ``Large scale distributed
  deep networks,'' \emph{Advances in neural information processing systems},
  vol.~25, 2012.

\bibitem{bianchini2021theoretical}
M.~Bianchini, P.~Frasconi, M.~Gori, M.~Maggini \emph{et~al.}, ``Optimal
  learning in artificial neural networks: A theoretical view,'' \emph{Neural
  network systems techniques and applications}, vol. 143, pp. 1--51, 1998.

\bibitem{he2016deep}
K.~He, X.~Zhang, S.~Ren, and J.~Sun, ``Deep residual learning for image
  recognition,'' in \emph{Proceedings of the IEEE conference on computer vision
  and pattern recognition}, 2016, pp. 770--778.

\bibitem{Shaukat2020}
Z.~Shaukat, S.~Ali, Q.~U.~A. Farooq, C.~Xiao, S.~Sahiba, and A.~Ditta,
  ``Cloud-based efficient scheme for handwritten digit recognition,''
  \emph{Multimedia Tools and Applications}, vol.~79, pp. 29\,537--29\,549,
  2020.

\bibitem{herculano2011scaling}
S.~Herculano-Houzel, ``Scaling of brain metabolism with a fixed energy budget
  per neuron: Implications for neuronal activity, plasticity, and evolution,''
  \emph{Proceedings of the National Academy of Sciences}, vol. 108, no.~10, pp.
  4230--4235, 2011.

\bibitem{raichle2002brainenergy}
M.~E. Raichle and D.~A. Gusnard, ``Appraising the brain’s energy budget,''
  \emph{Proceedings of the National Academy of Sciences}, vol.~99, no.~16, pp.
  10\,237--10\,239, 2002.

\bibitem{thorpe1996visualprocessing}
S.~Thorpe, D.~Fize, and C.~Marlot, ``Speed of processing in the human visual
  system,'' \emph{Nature}, vol. 381, no. 6582, pp. 520--522, 1996.

\bibitem{aghion2020labor}
\BIBentryALTinterwordspacing
P.~Aghion, C.~Antonin, S.~Bunel, and X.~Jaravel, ``What are the labor and
  product market effects of automation? new evidence from france,'' 2020.
  [Online]. Available:
  \url{https://scholar.harvard.edu/files/aghion/files/what_are_the_labor_and_product_market_effects_of_automation_jan2020.pdf}
\BIBentrySTDinterwordspacing

\end{thebibliography}
